\documentstyle[twoside,epsfig]{article}

\catcode`\@=11
\long\def\@makefntext#1{
\protect\noindent \hbox to 3.2pt {\hskip-.9pt
$^{{\eightrm\@thefnmark}}$\hfil}#1\hfill}               

\def\@makefnmark{\hbox to 0pt{$^{\@thefnmark}$\hss}}    

\def\ps@myheadings{\let\@mkboth\@gobbletwo
\def\@oddhead{\hbox{}
\rightmark\hfil\eightrm\thepage}
\def\@oddfoot{}\def\@evenhead{\eightrm\thepage\hfil
\leftmark\hbox{}}\def\@evenfoot{}
\def\sectionmark##1{}\def\subsectionmark##1{}}
\input{ijmpc1.sty}

\def\qed{\hbox{${\vcenter{\vbox{                        
   \hrule height 0.4pt\hbox{\vrule width 0.4pt height 6pt
   \kern5pt\vrule width 0.4pt}\hrule height 0.4pt}}}$}}

\def\bsc{{\sc a\kern-6.4pt\sc a\kern-6.4pt\sc a}}       
\def\bflatex{\bf L\kern-.30em\raise.3ex\hbox{\bsc}\kern-.14em
T\kern-.1667em\lower.7ex\hbox{E}\kern-.125em X}

\newcommand{\be}{\begin{equation}}
\newcommand{\ee}{\end{equation}}
\newcommand{\R}{{\cal R}}

\begin{document}

\runninghead{Generalized Deterministic Traffic Rules}
{Generalized Deterministic Traffic Rules}

\normalsize\textlineskip
\thispagestyle{empty}
\setcounter{page}{1}

\copyrightheading{}                     

\vspace*{0.88truein}

\fpage{1}
\centerline{\bf GENERALIZED DETERMINISTIC TRAFFIC RULES}
\vspace*{0.37truein}
\centerline{\footnotesize HENRYK FUK\'{S}}
\vspace*{0.015truein}
\centerline{\footnotesize\it Department of Physics, University of Illinois
at Chicago}
\baselineskip=10pt
\centerline{\footnotesize\it Chicago, IL , 60607-7059, USA}
\vspace*{10pt}
\centerline{\normalsize and}
\vspace*{10pt}
\centerline{\footnotesize
NINO BOCCARA\footnote{also at Department of Physics, University of
Illinois at Chicago,
 Chicago, IL 60607-7059}}
\vspace*{0.015truein}
\centerline{\footnotesize\it DRECAM/SPEC, CE Saclay}
\baselineskip=10pt
\centerline{\footnotesize\it  91191 Gif sur Yvette, France}
\vspace*{0.225truein}

\vspace*{0.21truein}
\abstracts{We study a family of deterministic models for highway traffic flow
which generalize cellular automaton rule 184. This family is
parametrized by the speed limit $m$ and another parameter $k$ that
represents a ``degree of aggressiveness'' in driving, strictly related
to the distance between two consecutive cars. We compare two driving
strategies with identical maximum throughput: ``conservative'' driving
with high speed limit and ``aggressive'' driving with low speed limit.
Those two strategies are evaluated in terms of accident probability.
We also discuss fundamental diagrams of generalized traffic rules and
examine limitations of maximum achievable throughput. Possible
modifications of the model are considered.}{}{}



\vspace*{1pt}\textlineskip      
\section{Introduction}
Transport phenomena in complex systems, in particular models of
highway traffic flow, attracted much attention in recent years. Much
of the effort was concentrated on discrete stochastic models of
traffic flow, first proposed by  Nagel and
Schreckenberg\cite{Nagel92}, and subsequently studied  by many other
authors using a variety of
techniques\cite{Schadschneider93,Vilar94,Nagel95,Schreckenberg95}. In
what follows, we shall study a family of purely deterministic traffic
models. The only randomness comes from the fact that the initial
configuration of cars is chosen at random. This family of models
represents various driving strategies, either chosen by drivers
(distance between cars) or externally imposed (such as the speed
limit). Our ``artificial highway'' consists of an array of $L$ cells.
Each cell is either occupied by a single car or empty. Cars can move
only to the right, and we assume periodic boundary conditions. Time is
discrete. At each time step, each driver moves his car according to
some specified rule. The evolution is synchronous, that is, all cars
move at the same time. In the simplest model of the family, car at
site $i$ can either move to site $i+1$ this site  is empty, or not
move if site $i+1$ is occupied. Thus, the state of a given cell $i$
depends only on cells $i-1$, $i$ and $i+1$. This model is equivalent
to cellular automaton\cite{Wolfram94} rule 184  if the state of an
occupied site is 1, whereas the state of an empty site is $0$. Under
this rule, which rule table is
\begin{eqnarray*}
& & 000 \rightarrow 0,\, 001 \rightarrow 0,\, 010 \rightarrow 0,\,
011 \rightarrow 1, \\ & & 100 \rightarrow 1,\, 101 \rightarrow 1,\,
110 \rightarrow 0,\, 111 \rightarrow 1,
\end{eqnarray*}
the density of 1's is conserved,  meaning that the number of ``cars''
does not change with time. Moreover, each car can move at most one
site to the right during one time step, so the ``speed limit'' for
this model is $m=1$. For a lattice of length $L$, the average speed
(sum of all speeds divided by the number of all cars) is, therefore,
always less or equal to $m=1$.

\begin{figure}
\begin{center}
\epsfxsize=4.5in \epsfbox{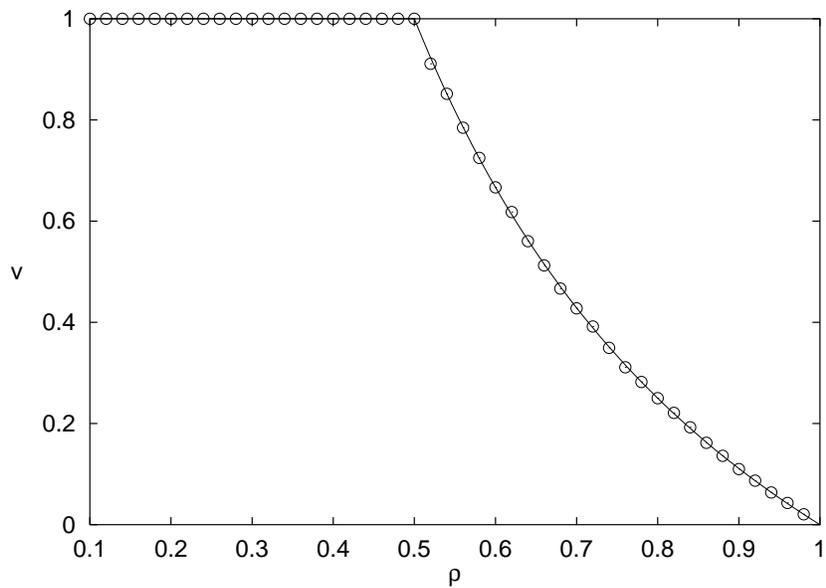}
\end{center}
\caption{Average speed of cars as a function of density for rule
184. Solid line represents theoretical $\overline{v}(\rho)$ curve.}
\label{r184}
\end{figure}

Let us now assume that we start from a random configuration of density
$\rho$, and that cars move according to rule 184. It has been recently
proved\cite{Fuks97} that for large $L$ and $t$, the average speed
$\overline{v}_t$ at time $t$ equals
\begin{equation}
\overline{v}_t = \left\{ \begin{array}{ll}
\Theta(\rho,t)  & \mbox{if $\rho<\frac{1}{2}$}, \\ [0.5em]
\displaystyle{ \frac{1-\rho}{\rho}}\Theta(\rho,t) &
\mbox{otherwise},
\end{array}
\right.
\end{equation}
where
\be
\Theta(\rho,t)=1-\frac{[4 \rho (1-\rho)]^t}{\sqrt{\pi t}}.
\ee
Since $\lim_{t \rightarrow \infty} \Theta(\rho,t)=1$, the average
speed in the long time limit $\overline{v}_\infty = 1$ when the car
density $\rho$ is less than $1/2$ and $\overline{v}_\infty=(1-\rho)/
\rho$ otherwise. In statistical physics terminology, the system
exhibits a second order kinetic phase transition, where $\rho$ is the
control parameter, and $\overline{v}$ the order parameter. The
critical point is at exactly $\rho=1/2$, and at the critical point
$\overline{v}_t$ approaches its stationary value as $t^{-1/2}$. Away
from the critical point, the approach is exponential, and it slows
down as$\rho$ comes closer to $1/2$. In fact, when the lattice is
finite (as in the real life), the behavior of this model is not
significantly different from the $L=\infty$ case. It is sufficient to
perform $L/2$ iterations in order to reach the stationary
state\cite{Fuks97}. These considerations are illustrated in Figure
\ref{r184}, representing the average car
 speed versus car density after 500 iterations of a 1000-site
lattice. Numerical simulations are compared to the theoretical
expression of $\overline{v}_\infty$ given by
\begin{equation}
\overline{v}_\infty = \left\{ \begin{array}{ll}
 1  & \mbox{if $\rho<\frac{1}{2}$}, \\ [0.5em]
 \displaystyle{\frac{1-\rho}{\rho}}    & \mbox{otherwise}.
\end{array}
\right.
\end{equation}
It is straightforward to generalize rule 184 to higher velocities. Let
us consider two cars $A$ and $B$ such that $A$ follows $B$, and denote
by $g$ the gap between them. Then, at the next time step, car $A$
moves $g$ sites to the right if $g \leq m$, and $m$ sites otherwise.
Such a rule will be referred to as $\R_{m,1}$, where $m$ denotes the
speed limit. With this notation, $\R_{1,1}$, represents rule 184. The
reason for the additional subscript ``1'' will become clear later. In
this case, the phase transition occurs at $\rho_c=1/(1+m)$. In the
free moving phase, i.e., below the critical point, all cars are moving
with maximum velocity. In the ``jammed   phase'' (above $\rho_c$), the
average speed is equal to $(1-\rho)/ \rho$, just as for rule 184.
\begin{figure}
\begin{center}
\epsfxsize=4.5in \epsfbox{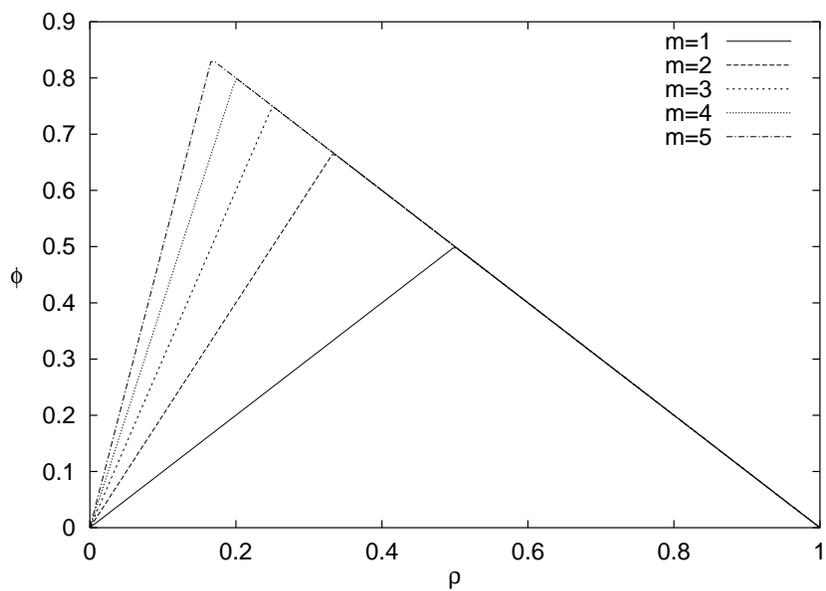}
\end{center}
\caption{Fundamental diagram for $\R_{m,1}$ for several different
values of $m$. Vertical axis represents the flow $\phi=\rho
\overline{v}$.}
\label{flowa}
\end{figure}
The plot of the flow $\phi=\rho \overline{v}$ versus $\rho$ for
several different values of $m$, called the fundamental
diagram, is shown in Figure \ref{flowa}. It is clear that when the
speed limit $m$ increases, the total throughput of the highway
increases.

\section{Monitored Traffic}
How can we increase the fluidity of the flow without increasing the
speed limit? It is obvious that the source of inefficiency in rule
184 is the ``defensive attitude'' of the drivers. Consider, for
example, the following configuration:
\vspace*{10pt}
\begin{center}
\begin{tabular}{|c|c|c|c|c|c|c|c|c|c|} \hline
$\cdots$&0&A&B&0&C&0&0&0&$\cdots$\\ \hline
$\cdots$&0&A&0&B&0&C&0&0&$\cdots$\\ \hline
\end{tabular}
\end{center}
\vspace*{10pt}
The first line represents locations of cars A, B, C at time $t$, and
the second line their location at time $t+1$. Zeros represent empty
sites. Since driver A doesn't know whether car B is going to move or
not, it is safer for him not to move. If he could see further than
just one site forward, he could predict that car B will move forward,
and that the site in front of him will be vacant at the next time
step.

\begin{figure}
\begin{center}
\epsfxsize=4.5in \epsfbox{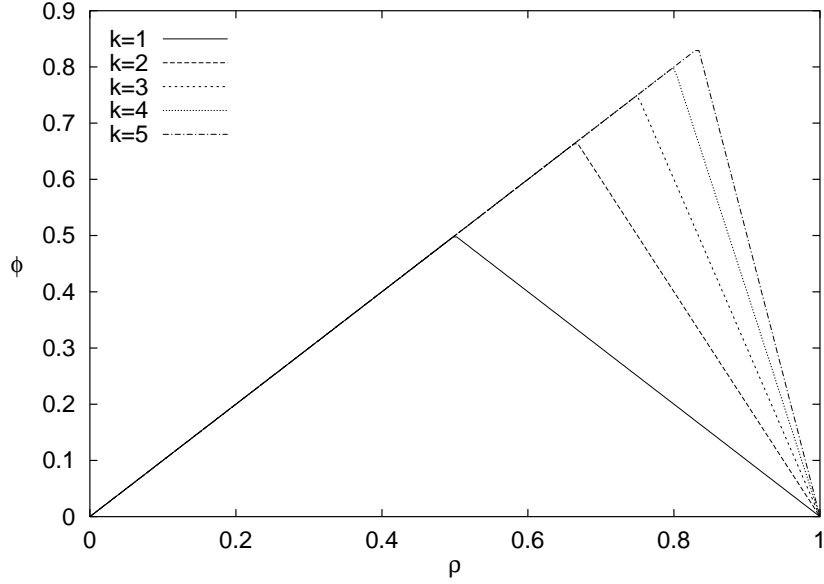}
\end{center}
\caption{Fundamental diagram for $\R_{1,k}$ for several different
values of $k$.}
\label{flowb}
\end{figure}

A simple rule which incorporates such a ``prediction'' mechanism can
be constructed in a following way. Let us say that a driver at site
$i$ first checks whether in a block of $k$ sites directly in front of
him---i.e., from site $i+1$ to site $i+k$---at least one site is empty
(we will say that to be able to get this information cars have to be
``monitored''). If it is, he moves his car by one site to the right,
even if site $i+1$ is occupied since he knows that the car at site
$i+1$ will move because all drivers follow the same rule. If all sites
from $i+1$ to $i+k$ are occupied, the car at site $i$ does not move.
Such a his will be denoted by $R_{1,k}$. For $k=2$, the configuration
discussed before will evolve as
\vspace*{10pt}
\begin{center}
\begin{tabular}{|c|c|c|c|c|c|c|c|c|c|} \hline
$\cdots$&0&A&B&0&C&0&0&0&$\cdots$\\ \hline
$\cdots$&0&0&A&B&0&C&0&0&$\cdots$\\ \hline
\end{tabular}
\end{center}
\vspace*{10pt}
As we see, cars A and B now move as one block, as long as the site
in front of car B is empty. Such ``blocking'' significantly
improves fluidity, as the fundamental diagram  in Figure
\ref{flowb} demonstrates.
We immediately notice that fundamental diagrams corresponding to rules
$\R_{1,n}$ and $\R_{n,1}$ are mirror images of each other with respect
to the $\rho=0.5$ line. Indeed $\R_{1,n}$ and  $\R_{n,1}$ are closely
related: If cars are moving to the right according to rule $\R_{1,n}$,
then empty sites are moving to the left according to rule $\R_{n,1}$.
The critical density, which corresponds to a periodic configuration
with a period consisting of $k$ cars followed by one empty site equals
$\rho_c=k/(k+1)$. Above $\rho_c$, a similar relationship holds, i.e.,
$\rho_c=k/(k+\overline{v})$. Hence
\begin{equation}
\label{vinftybr}
\overline{v}_\infty = \left\{ \begin{array}{ll}
 1  & \mbox{if $\rho<k/(k+1)$}, \\
 k(1-\rho)/ \rho    & \mbox{otherwise}.
\end{array}
\right.
\end{equation}

\section{Accidents}
Rules $\R_{n,1}$ and $\R_{1,n}$ may be considered as two complementary
driving strategies. Drivers obeying rule $\R_{n,1}$ drive with higher
speed, but they keep the distance to the preceding car larger. Those
who comply with rule $\R_{1,n}$ drive more slowly, however they keep a
smaller distance between consecutive cars. Since the maximum possible
flow for both rules is the same (it equals $n/(n+1)$), one could ask
which rule is better in term of driving safety. Of course, if all
drivers follow the same rule, either $\R_{n,1}$ or $\R_{1,n}$, there
is no problem , since no accident can occur. But, in the real world,
some drivers are careless, have slower reaction time, defective brakes
etc., and accidents do occur. On the single-lane highway, only one
type of accident may occur: a car bumps into the preceding car which
abruptly decreased its velocity. Thus, we can say that cars which
decrease their velocity are potential causes of accidents. In our
model, such potentially dangerous cars can be identified as cars
which, at time $t$, have a smaller velocity that at time $t-1$ (by
definition the velocity at time $t$ is $x(t)-x(t-1)$, where $x(t)$ is
a position of a given car at time $t$). A simple mean-field estimate
of a number of such cars can be carried out if we neglect time
correlations between velocities, i.e., when we assume that $v_{t-1}$
and $v_t$ are not correlated. For rule $\R_{m,1}$, the probability
that a given car is slowing down will then be given by
\be
P(v_t<v_{t-1})=\sum_{i=1}^m \sum_{j=0}^{i-1} P(v_t=j) P(v_{t-i}=i),
\ee
where $P(v_t=j)$ denotes the probability that a car has velocity
$j$ at time $t$. Let us further assume that the length of the
lattice is $L$, the number of cars is $N$, and the number of cars
which have velocity $i$ is $N_i$. In the stationary state $N_i$ is
time-invariant, thus we can write
\be
P(v_t<v_{t-1})=\sum_{i=1}^m \sum_{j=0}^{i-1}n_j n_i,
\ee
where $n_i=N_i/N$. The values of $n_i$'s for $\R_{m,1}$ are given
by\cite{Fukui96c}
\begin{eqnarray}
n_k&=&n_0 (1-n_0)^k  \mbox{ if } k<m \\ \nonumber
n_m&=&(1-n_0)^m
\end{eqnarray}
Using these expressions, we obtain
\begin{eqnarray*}
P(v_t&<&v_{t-1})=
 \sum_{j=0}^{i-1}n_j n_m +
\sum_{i=1}^{m-1} \sum_{j=0}^{i-1}n_j n_i= \\
&=&n_0(1-n_0)^m \sum_{j=0}^{m-1}(1-n_0)^j + \\
&+&n_0^2 \sum_{i=1}^m \sum_{j=0}^{i-1} (1-n_0)^j (1-n_0)^i,
\end{eqnarray*}
and after computing all sums,
\begin{eqnarray*}
\lefteqn{P(v_t<v_{t-1})=1-n_0 -(1-n_0)^{2m} +} \hspace{1.2cm} \\
&& \mbox{} +  \displaystyle{\frac{(1-n_0)^{2m}+2n_0-1-n_0^2}{2-n_0}
}.
\end{eqnarray*}
As shown by Fukui and Ishibashi\cite{Fukui96c}, when $\rho>\rho_c$,
$n_0$ is a solution of the following nonlinear equation:
\be
\frac{(1-n_0)[1-(1-n_0)^m]}{n_0}=\frac{1}{\rho} -1,
\ee
If, for example, $m=2$, this equation can be solved and $n_0$ is
given by
\be
n_0=\frac{3}{2} - \frac{1}{2}\sqrt{\frac{4-3\rho}{\rho}}.
\ee
The fraction of cars which are slowing down equals
\be
\label{fscr2}
P(v_t<v_{t-1})=\frac{3\rho^2 - 5\rho -2}{2 \rho^2} +
\frac{3+\rho}{2 \rho} \sqrt{\frac{4-3\rho}{\rho}}.
\ee
Below $\rho_c$, $P(v_t<v_{t-1})$ is of course zero, since all cars
are moving with constant speed $m$.

For $\R_{1,k}$, the mean-field approximation is much simpler,
because all cars are either stopped or moving with speed 1 and
$\overline{v}=0\times n_0+1\times n_1=n_1$,
thus
\be
P(v_t<v_{t-1})=n_0 n_1 =(1-n_1) n_1= (1-\overline{v})\overline{v}.
\ee
For $k=2$, equation (\ref{vinftybr}) yields
$\overline{v}=2(1-\rho)/ \rho$ (above the critical density),
therefore
\be
\label{fscbr2}
P(v_t<v_{t-1})=\frac{1(1-\rho)(3\rho-2)}{\rho}.
\ee
As before, below $\rho_c$ the fraction of cars which are slowing
down is zero.

\begin{figure}
\begin{center}
\epsfxsize=4.5in \epsfbox{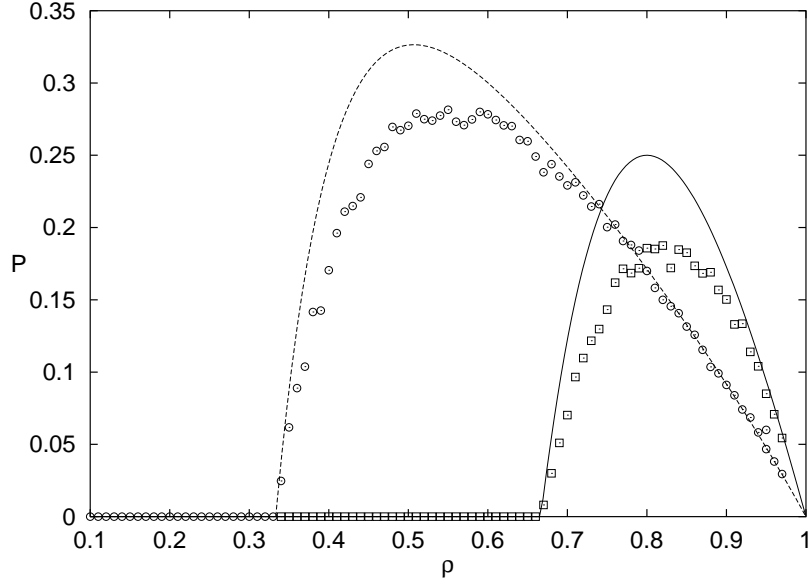}
\end{center}
\caption{Fraction of cars which are slowing down for $\R_{1,2}$
($\Box$) and $\R_{2,1}$ ($\circ$). Solid and dashed lines represent
the mean-field approximation for, respectively, rules $\R_{1,2}$
and $\R_{2,1}$.}
\label{slowing}
\end{figure}

The fraction of slowing cars obtained from computer simulations for
both $\R_{2,1}$ and $\R_{1,2}$ is shown in Figure \ref{slowing},
together with mean-field approximation curves given by equations
(\ref{fscr2}) and (\ref{fscbr2}). Although mean-field predictions
overestimate $P(v_t<v_{t-1})$, they are not very far from
``experimental'' results. One feature, however, is apparent: accident
probability, which should be proportional to $P(v_t<v_{t-1})$, is much
higher for $\R_{2,1}$ than for $\R_{1,2}$ if the density of cars is
below 0.8. Above 0.8, rule $\R_{1,2}$ become more ``dangerous'',
although the difference between rules diminishes as $\rho$ approaches
1. It is also remarkable that in the case of rule $\R_{1,2}$ we have
no accidents up to $\rho=2/3$, while rule $\R_{2,1}$ becomes very
dangerous quite fast, being worst at approximately $\rho=0.5$.
\newpage
\section{Generalization}
\begin{figure}
\begin{center}
\epsfxsize=4.5in \epsfbox{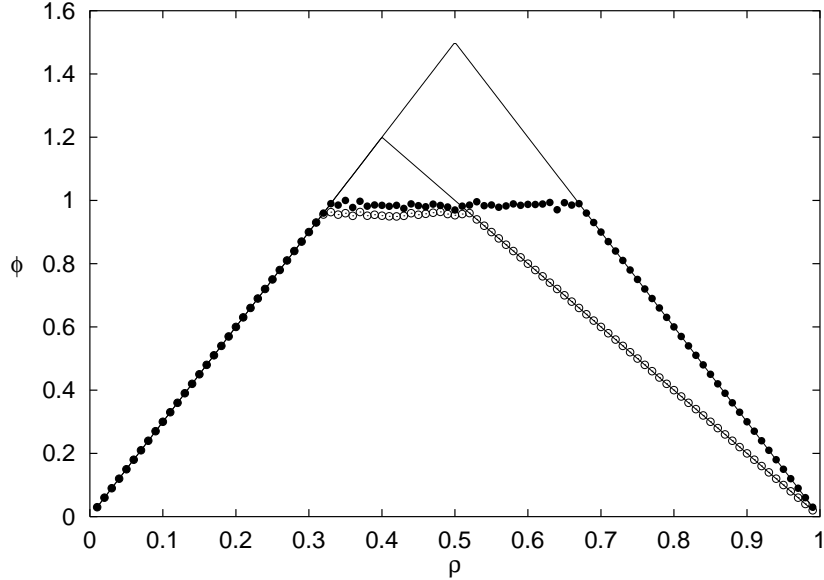}
\end{center}
\caption{Fundamental diagram for $\R_{3,3}$ ($\bullet$) and
$\R_{3,2}$ ($\circ$). Solid lines represent theoretical flow
obtained from eq. (\ref{gentheoryf}). }
\label{genflux}
\end{figure}

\begin{figure}[h]
\begin{center}
\epsfxsize=4.5in \epsfbox{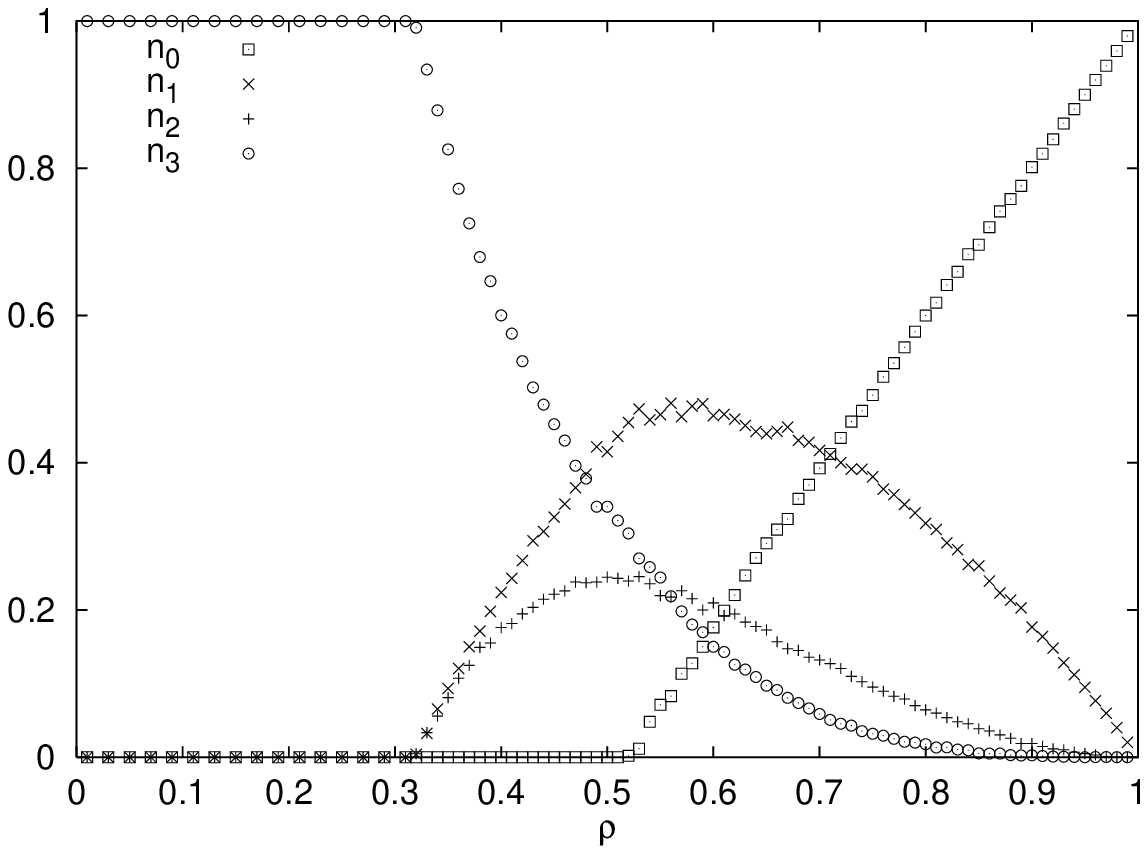}
\end{center}
\caption{Velocity spectrum for $\R_{3,2}$. Spectrum computed after
$5000$ iterations, lattice size $L=2000$.}
\label{spectrum}
\end{figure}

Let us now construct a general rule which allows both higher speed
limit $m$ and blocking of $k$-th order, with $m$ and $k$ both
larger than 1. Consider a car located at site $i$. Its driver first
locates the nearest gap  (cluster of empty sites) in front of him
whose length is $g$. If the first empty site (i.e., the first site
belonging to the mentioned gap) is farther than $i+k$, the car does
not move, otherwise, it moves to site $i+v$ site, where $v=\min
(g,m)$. We will refer to this rule as $\R_{m,k}$. We could now
expect that, as before, the ``perfect configuration'' is a periodic
sequence whose period consists of $k$ cars followed by $m$ empty
sites. The critical density is then $\rho_c=k/(k+m)$, and this
should hold for any $\rho$ larger than $\rho_c$, i.e.,
$\rho=k/(k+\overline{v})$. Below $\rho_c$, we expect
$\overline{v}=m$. For the average velocity, this leads to the
following expression:
\be
\label{gentheoryv}
v = \left\{ \begin{array}{ll}
 m  & \mbox{if $\rho < \rho_c$}, \\
 k(1-\rho)/ \rho  & \mbox{otherwise}.
\end{array}
\right.
\ee
Similarly, the flow is given by
\be
\label{gentheoryf}
\phi = \left\{ \begin{array}{ll}
 \rho m  & \mbox{if $\rho < \rho_c$}, \\
 (1-\rho)k  & \mbox{otherwise}.
\end{array}
\right.
\ee
Unfortunately, this simple reasoning has some flaws. Although in the
vicinity of $\rho=0$ and $\rho=1$ equations (\ref{gentheoryv}) and
(\ref{gentheoryf}) describe the behavior of $\R_{m,k}$ correctly, they
fail to give a correct description in the intermediate region. This is
clearly illustrated by the fundamental diagram, which, according to
(\ref{gentheoryf}), should be always tent-shaped, with a peak at
$\rho_c=k/(k+m)$. In practice, however, the shape is quite different,
as shown in Figure \ref{genflux}.
There is no single peak at $\rho_c$, but instead, the flow stays at
its maximum value over an extended interval of $\rho$ values, and the
fundamental diagram looks like ``a tent with a flat roof''. This means
that, unlike rules $\R_{1,k}$ and $\R_{m,1}$, rule $\R_{m,k}$ for
$m,k>1$ does not exhibit a phase transition at $\rho_c$. This is due
to the fact that the ``perfect configuration'' with period consisting
of $k$ cars followed by $m$ empty sites is no longer stable, like it
was the case for rule $\R_{1,k}$ or $\R_{m,1}$. In fact, the phase
transition does occur, but it is, in a sense, spread over an interval.
To be more precise, let us first note that $\overline{v}$ is a linear
combination of $n_i$'s. Consequently, if, as functions of $\rho$, any
of the $n_i$ has a discontinuous derivative, this will show up in
$\overline{v}$. Figure \ref{spectrum} shows an example of such a
``velocity spectrum'', i.e., plots of $n_i(\rho)$ for $i=0,1,2,3$. We
can see that derivatives of all $n_i$'s have discontinuities at some
point, but they occur at different values of $\rho$. The first
transition occurs at $\rho=0.33$ ($n_3=1 \rightarrow n_3 \neq1$,
$n_1=0 \rightarrow n_1 \neq 0$ and $n_2=0 \rightarrow n_2 \neq0$ ),
whereas at $\rho \approx 0.51$ we have another transition $n_0=0
\rightarrow n_0 \neq0$. For rule $\R_{m,1}$, all these transitions
occurred at the same point, and we consequently observed a single
transition in $\overline{v}$.

Another interesting feature of the fundamental diagram of
$\R_{m,k}$ is the maximum flow value. From (\ref{gentheoryf}) one
would expect $\phi_{max}=km/(k+m)$, which could be as large as we
want if we choose the right $m$ and $k$ values. Figure
\ref{genflux}, however, clearly shows that this is not the case:
there seems to be a cutoff at $\phi$ slightly below $1$. In fact,
simulations performed with a wide array of $k$ and $m$ values
suggest that the flow can never be larger that $1$, regardless of
$k$ and $m$. For large $k$ and $m$, the cutoff occurs almost
exactly at $1$, so the flow is well approximated by the piecewise
linear function
\be
\phi = \left\{ \begin{array}{ll}
 \rho m     & \mbox{if $\rho \leq 1/m$}, \\
   1        & \mbox{if $1/m < \rho < (k-1)/k$}, \\
 (1-\rho)k  & \mbox{if $\rho \geq (k-1)/k$}.
\end{array}
\right.
\ee
``Large'' means equal or larger than $3$, as even for $m=k=3$ the
above formula is fairly accurate (the cutoff occurs at 0.98,
instead of 1). In any case, $\phi=1$ is the absolute maximum flow
in the stationary state. It is not possible to make it larger by
increasing either $k$ or$m$.

\section{Possible Modifications}
Driving strategies based on $\R_{m,k}$ are, of course, not the only
possible ways of increasing traffic fluidity. In the case of rule
$\R_{m,k}$, the decision of a driver where to move the car at the
next time step is based only on the size of the nearest gap
$g_{near}$, located within the $k+m-1$ sites in front of the car.
Instead of considering the nearest gap, drivers could base their
decisions on the largest gap. More precisely, let us say that
$g_{max}$ is the largest continuous block of empty sites, entirely
located between sites $i$ and $i+k+m$. If, for example, $m=3$,
$k=7$, consider the configuration $\ldots
\underline{1}0100110000\ldots$, the largest continuous block of
zeros within $m+k-1=9$ sites in front of the underlined car has a
length $g_{max}=3$. We now define rule $\hat{\R}_{m,k}$ such that
drivers move their cars by $\min(g_{max},m)$ sites to the right.
Since $g_{max}\geq g_{near}$, we can expect that the average speed
(or flow) for $\hat{\R}_{m,k}$ should be larger or equal than the
average speed (or flow) for $\R_{m,k}$.
\begin{figure}
\begin{center}
\epsfxsize=4.5in \epsfbox{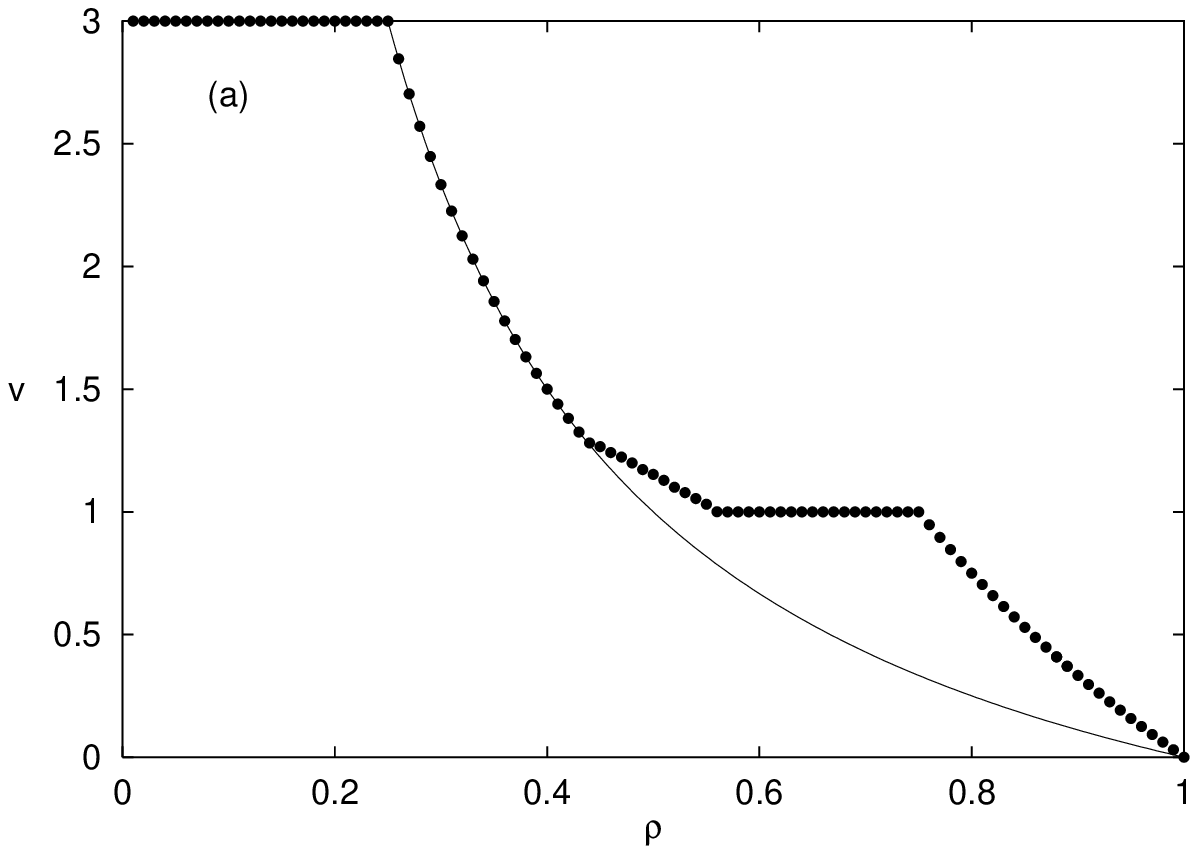}
\epsfxsize=4.5in \epsfbox{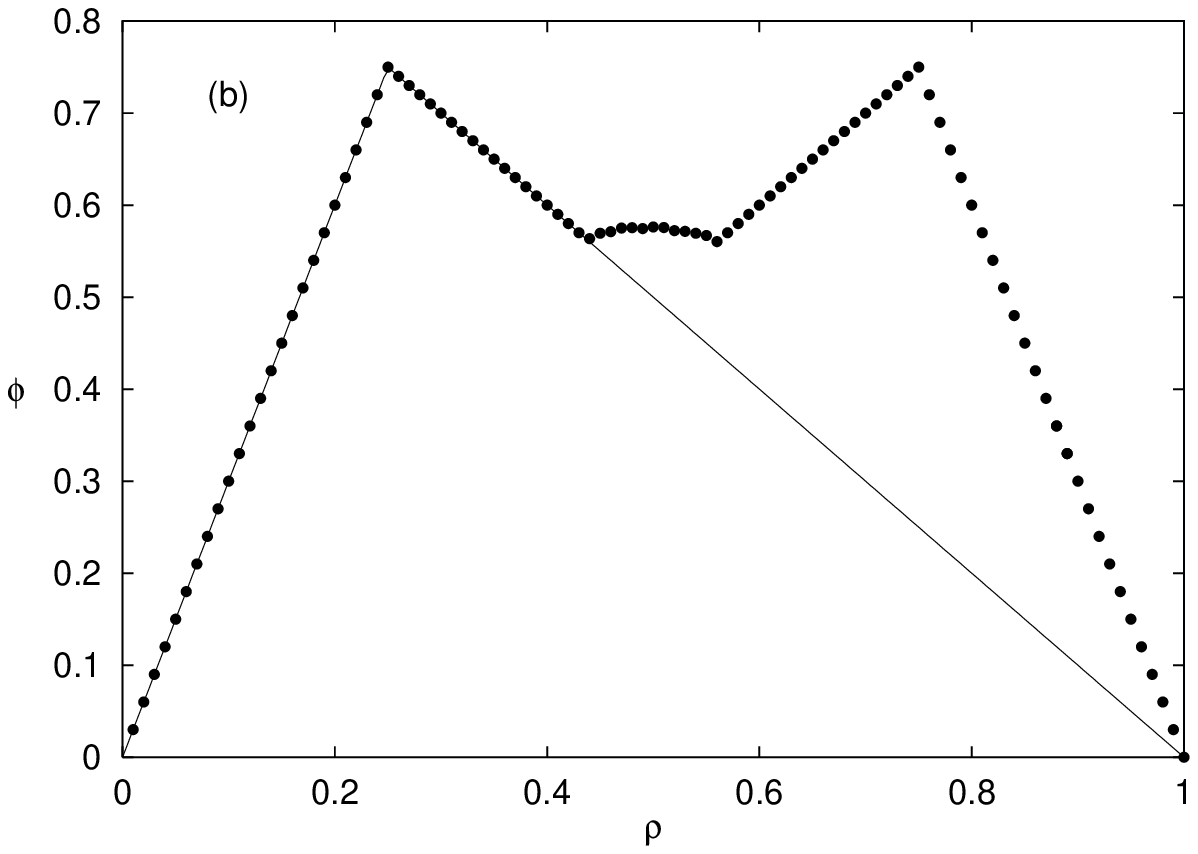}
\end{center}
\caption{(a) Plot of average velocity $\overline{v}$ as a function
of density $\rho$ for $\hat{\R}_{3,1}$ ($\bullet$) and $\R_{3,1}$
(solid line). (b) Fundamental diagram of $\hat{\R}_{3,1}$
($\bullet$) and $\R_{3,1}$ (solid line).}
\label{hat31}
\end{figure}
To demonstrate that this is indeed the case, let us compare rules
$\hat{\R}_{3,1}$ and $\R_{3,1}$. Figure \ref{hat31}a shows that up to
$\rho \approx 0.43$, $\hat{\R}_{3,1}$ behaves exactly like $\R_{3,1}$.
When the density increases beyond this value, additional phase
transitions occur. While rule $\R_{3,1}$ exhibits just a single phase
transition at $\rho=0.25$, $\hat{\R}_{3,1}$ exhibits an entire
cascade, at $\rho=0.25$, $0.43$, $0.57$, and $0.75$. Interestingly,
looking at the fundamental diagram (Figure \ref{hat31}b) we can see
that it is fully symmetric with respect to $\rho=0.5$. This means that
the second peak has the same shape as the peak in fundamental diagram
of $R_{3,1}$ (as a consequence of the duality between $\R_{3,1}$ and
$\R_{1,3}$). Therefore, $\hat{\R}_{3,1}$ can be considered as a sort
of the ``mix'' of the aforementioned rules,
\be
\hat{\R}_{3,1} \approx \left\{ \begin{array}{ll}
  \R_{3,1}     & \mbox{if $\rho <0.5$}, \\
  \R_{1,3}     & \mbox{if $\rho >0.5$}.
\end{array}
\right.
\ee
The $\approx$ sign reflects the fact that this relation is not true
in the vicinity of $0.5$.

Another interesting detail which should be mentioned pertains to
the maximum flow value. As in the $\R_{m,k}$ case, we investigated
the fundamental diagram of $\hat{\R}_{m,k}$ for many different $m$
and $k$ values, and it seems that also in this case $\phi$ cannot
exceed 1. Nevertheless, we found no simple explanation of this
seemingly general feature.

\end{document}